\documentclass[conference]{IEEEtran}
\usepackage{lipsum}
\usepackage{fancyhdr}
\usepackage{url}
\usepackage{fancyheadings}
\fancypagestyle{firstpage}{%
  \fancyhf{}
  
 \fancyfoot[C]{\textit{\small{IEEE International Conference on Cyber Security and Protection of Digital Services (Cyber Security 2017), June 19-20, 2017, London, UK.}}}
}

\ifCLASSINFOpdf
\else
\fi
\usepackage{graphicx}
\usepackage{pifont}
\usepackage{caption}
\usepackage{cite}
\usepackage{enumitem}
\usepackage{threeparttable}
\usepackage{amsmath}
\usepackage[algo2e]{algorithm2e}

\usepackage{algorithmic}
\begin{document}
%
\title{Continuous Implicit Authentication for Mobile Devices based on Adaptive Neuro-Fuzzy Inference System }




\author{\IEEEauthorblockN{Feng Yao\IEEEauthorrefmark{1},
Suleiman Y. Yerima\IEEEauthorrefmark{2},
BooJoong Kang\IEEEauthorrefmark{3} and
Sakir Sezer\IEEEauthorrefmark{4}}
\IEEEauthorblockA{\IEEEauthorrefmark{1} Email: fyao02@qub.ac.uk  \IEEEauthorrefmark{2} Email: s.yerima@qub.ac.uk}
\IEEEauthorblockA{\IEEEauthorrefmark{3} Email: B.Kang@qub.ac.uk  \IEEEauthorrefmark{4} Email: s.sezer@ecit.qub.ac.uk}
\IEEEauthorblockA{Centre for Secure Information Technologies (CSIT)\\Queen's University Belfast, Belfast, BT3 9DT, Northern Ireland, UK}}


\maketitle

\thispagestyle{firstpage}

\begin{abstract}
As mobile devices have become indispensable in modern life, mobile security is becoming much more important. Traditional password or PIN-like point-of-entry security measures score low on usability and are vulnerable to brute force and other types of attacks. In order to improve mobile security, an adaptive neuro-fuzzy inference system(ANFIS)-based implicit authentication system is proposed in this paper to provide authentication in a continuous and transparent manner. To illustrate the applicability and capability of ANFIS in our implicit authentication system, experiments were conducted on behavioural data collected for up to 12 weeks from different Android users. The ability of the ANFIS-based system to detect an adversary is also tested with scenarios involving an attacker with varying levels of knowledge. The results demonstrate that ANFIS is a feasible and efficient approach for implicit authentication with an average of 95\% user recognition rate. Moreover, the use of ANFIS-based system for implicit authentication significantly reduces manual tuning and configuration tasks due to its self-learning capability.
\end{abstract}


%
\IEEEpeerreviewmaketitle

\section{Introduction}
Mobile devices are becoming more and more indispensable in our daily life. Technological advancements and the availability of high speed connections has made the mobile device a handy tool for managing day-to-day tasks involving sensitive data, monetary transactions etc. As a consequence, more threats are appearing and compromising security and privacy of millions of users' data worldwide. A survey~\cite{bib1} conducted by Kaspersky Consumer Security Risk shows that for every 6 mobile users, there is one who has suffered loss or theft. In order to mitigate the effect of loss or theft, novel efficient and intelligent access control techniques should be put in place while ensuring minimum impact on user experience.

PINs, passwords and pattern locks have long been used to provide mobile access control for simplicity and ease of use. However, these traditional point-of-entry authentication techniques have drawbacks since they only authenticate at the beginning of a session, after which users can access everything on the phone without being re-authenticated periodically. Moreover, weak passwords and PINS are vulnerable to brute force or dictionary attacks. Long and complicated passwords score low on usability, especially on mobile devices. Pattern lock can also be cracked within five attempts regardless of its complexity as stated in~\cite{bib2}. To make matters worse, nowadays most applications have a default setting to only require login credentials at first use, which put users' sensitive data at risk. Biometrics-based authentication techniques~\cite{bib3} such as fingerprint, iris or voice recognition are gaining popularity. However, they can still be compromised in some circumstances and require specific hardware support as well as higher computational resources.

As traditional authentication methods fall short of security and usability, implicit authentication (IA) is gaining attention as a complementary authentication method which is capable of authenticating a legitimate user and detecting imposters transparently and continuously without the explicit involvement of the user. 
According to a survey~\cite{bib4}, IA is widely accepted, and 73\% of the participants consider IA more secure. Recent researches~\cite{bib5,bib6,bib7,bib8,bib9,bib10,bib11,bib12,bib13,bib14,bib15} demonstrate the use of behavioural-based IA to provide continuous protection on mobile devices. Although the results are promising, maintaining a balance between accuracy, adaptiveness and practical feasibility is still an unresolved challenge. Hence, an ANFIS-based implicit authentication system is presented in this paper. By utilizing its self-learning capabilities, the ANFIS-based system is able to ascertain the unique behavioural pattern for every single user. Based on the perceived pattern, the ANFIS-based system is able to provide continuous and transparent authentication in real-time. Our proposed ANFIS-based IA system incorporates a time-window based profiling approach in order to enable granular and continuous re-authentication which also ensures timely adversary detection. The proposed system also utilizes anomaly based scoring, which together with an adaptively computed reference, provides real-time inputs for the ANFIS system. The anomaly based scoring system applies a ranking algorithm that maintains a sorted list of the most relevant events that are used in the computation of the anomaly-based scores.


In the rest of the paper, section~\ref{sec2} reviews background of ANFIS while section~\ref{sec3} describes our proposed authentication system. Section~\ref{sec4} presents the evaluation of the system using real data collected from different users over several weeks. Section~\ref{sec5} summarizes related work regarding behaviour-based implicit authentication. Section~\ref{sec6} concludes the paper and proposes directions for future work.

\section{Background of ANFIS}
\label{sec2}
ANFIS is a fuzzy inference system implemented in the framework of adaptive neural networks to facilitate learning and adaptation~\cite{bib16}. Neural network is not ideal for representing vagueness and uncertainty. As for fuzzy logic, the determination of fuzzy rules, input and output membership functions requires prior knowledge and cumbersome tuning process. However, by combining fuzzy logic and neural network technology together in ANFIS, it can fully take advantage of learning capabilities of neural network and imprecision handling of fuzzy logic to overcome these weaknesses. ANFIS has been applied in numerous areas including dissolved gas analysis~\cite{bib17}, identification and monitoring of brain-computer interface~\cite{bib18} and automatic gas tungsten arc welding~\cite{bib19}.

\subsection{Architecture of ANFIS}
\label{sec2.1}
The architecture of ANFIS is presented in Fig.~\ref{fig_anfis_architecture}. For simplicity, it is a first-order Sugeno fuzzy model with two inputs and two if-then rules.

\vspace{2.5 mm}
Rule 1: If ($x$ is $A_1$) and ($y$ is $B_1$) then $f_1 = p_1x+q_1y+r_1$

Rule 2: If ($x$ is $A_2$) and ($y$ is $B_2$) then $f_2 = p_2x+q_2y+r_2$

\vspace{2.5 mm}{

\begin{figure}
\centering
\includegraphics[width=1\columnwidth,height=4cm]{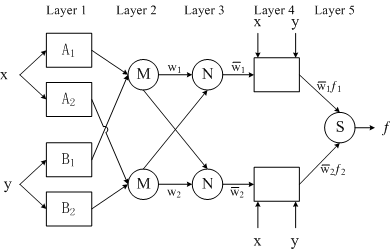}
\caption{ANFIS architecture for two-input Sugeno fuzzy model with two rules}
\label{fig_anfis_architecture}
\end{figure}
where $x$ and $y$ are inputs, $f_i$ are outputs, $A_i$ and $B_i$ are fuzzy sets. $p_i$, $q_i$ and $r_i$ are the consequent parameters which are updated during training process. 
In this figure, a square represents an adaptive node while a circle represents a fixed node. As you can see, all the nodes in layer 1 are adaptive nodes. Let the output of the $i^{th}$ node in layer $n$ be $O_i^n$, thus the outputs of layer 1 representing membership grade of the inputs can be represented as follows:
\begin{equation}\label{2.1.1}
\begin{cases}
O_{i}^{1} = \mu_{A_i}(x), & i=1,2 \\
O_{i}^{1} = \mu_{B_{i-2}}(y), & i=3,4 \\
\end{cases}
\end{equation}
where $\mu_{A_i}(x)$ and $\mu_{B_{i-2}}(y)$ can be any type of membership functions. For example, if sigmoid membership function is employed, $\mu_{A_i}(x)$ can be denoted as:
\begin{equation}\label{2.1.2}
\mu_{A_i}(x) = \frac{1}{1+e^{-c_i(x-b_i)}}
\end{equation}
where $c_i$, $b_i$ are premise parameters used to control the shape of membership functions.

In the second layer, every node is a fixed node which functions as a multiplier for incoming signals:
\begin{equation}\label{2.1.3}
O_{i}^{2} = w_i = \mu_{A_i}(x)\cdot \mu_{B_i}(y)
\end{equation}
Each node output represents the firing strength of a rule.

Every node in the third layer is a fixed node labeled with N, which indicates that they perform normalization to the firing strength from previous layer. The output of this layer is shown as follows:
\begin{equation}\label{2.1.4}
O_{i}^{3} = \bar{w}_i = \frac{w_i}{w_1+w_2}  \hspace{5mm}  i = 1, 2
\end{equation}

In the fourth layer, every node is an adaptive node, the output of this node is simply a product of normalized firing strength and first order polynomial:
\begin{equation}\label{2.1.5}
O_{i}^{4} = \bar{w}_if_i = \bar{w}_i(p_ix+q_iy+r_i)  \hspace{3mm}  i = 1, 2
\end{equation}

There is only one fixed node in fifth layer which performs the summation of all incoming signals. Hence, the output of fifth layer is given as:
\begin{equation}\label{2.1.6}
O_{i}^{5} = \sum_{i=1}^{2}\bar{w}_if_i = \frac{\sum_{i=1}^{2}w_if_i}{w_1+w_2} 
\end{equation}
\subsection{Learning algorithms of ANFIS}
\label{sec2.2}
As discussed in the previous section, premise parameters $\{c_i,b_i\}$ in layer 1 and consequent parameters $\{p_i,q_i,r_i\}$ in layer 4 are all modifiable, the purpose of the learning process of ANFIS is to tune all these parameters to make the ANFIS output match the training data. When the premise parameters are fixed, the final output can be represented as:
\begin{equation}\label{2.1.6}
\begin{aligned}
\text{$f$}& = \bar{w}_1f_1 + \bar{w}_2f_2 \\
&= \bar{w}_1(p_1x+q_1y+r_1) + \bar{w}_2(p_2x+q_2y+r_2)\\
&= (\bar{w}_1x)p_1 + (\bar{w}_1y)q_1 + (\bar{w}_1)r_1 + (\bar{w}_2x)p_2 \\
& \hspace{3mm}+(\bar{w}_2y)q_2 + (\bar{w}_2)r_2
\end{aligned}
\end{equation}

\begin{figure*}[ht]
\centering
\includegraphics{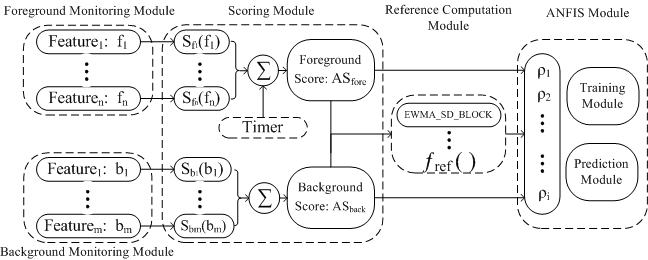}
\caption{ANFIS-based implicit authentication system diagram}
\label{fig_sbd}
\end{figure*}

which is a linear combination of consequent parameters. Thus, least square method can be used to compute the optimal value for consequent parameters. When the premise parameters are not fixed, the search space becomes larger and the convergence of the training becomes slower. The hybrid learning algorithm comprises of a forward pass and a backward pass. Once the optimal consequent parameters are found during the forward pass, the backward pass starts. In the backward pass, errors are propagated backward and premise parameters regarding membership function are updated by gradient descent. When the training is finished, both premise and consequent parameters are optimized and can be employed for prediction or classification. The combination of least square method and gradient descent has proven to be highly effective at training ANFIS models~\cite{bib16}.

\section{ANFIS-based Authentication System}
\label{sec3}

In this section, the structure of our system and functionalities of each component are described. The structure of ANFIS-based implicit authentication system is presented in Fig.~\ref{fig_sbd} which includes an activity monitoring module, a scoring module, a reference computation module and an ANFIS module. The learning capability of ANFIS allows for training of specific fuzzy inference model based on given input-output user data without manual interference. Therefore, this trained fuzzy model represents well the user's behaviour and can then be used to make final authentication decision to provide access control. The deployment of the system can be divided into two phases: training phase and deployment phase. During the training phase, we use generated input (scores, references) and output (classification labels) variables to train ANFIS module offline. When the training is completed the deployment phase starts, this enables the system to infer the authenticity of the current user behaviour by taking in real-time computed scores and references.

\subsection{Workflow of ANFIS-based authentication system}
\label{sec3.1}
The workflow of the proposed authentication system is illustrated in Fig.~\ref{fig_sbd}. It firstly models the user profile based on selected features that  are representative of individual behaviour. The features are divided into foreground and background activities. The background activities consists of those events that the user can not control e.g. receiving SMS and phone calls. The foreground activities on the other hand, consists of those the user can initiate e.g. making a phone call or browsing the Internet. The features used to score foreground and background behaviour can be represented as:
\begin{equation}\label{3.1.1}
Features: (f_1,f_2,f_3,...,f_n),(b_1,b_2,b_3,...,b_m)
\end{equation}

A time window-based profiling technique is adopted to adaptively compute the real-time foreground and background anomaly score to quantify the anomaly level of current user behaviour. Because user behaviour is dynamic, a generalized ranking algorithm is developed (see section III C) to maintain a record of the top n most recent or most relevant events for each feature (except the Screen On/Off feature). The values generated by the ranking algorithm within each feature category, are used in the anomaly score computation as described in section III C. The scoring functions to compute the anomaly score for each feature are represented as:
\begin{equation}\label{3.1.2}
Functions: (S_{f1},S_{f2},...,S_{fn}),(S_{b1},S_{b2},...,S_{bm})
\end{equation}
Because the foreground anomaly score $AS_{fore}$ is calculated incrementally based upon all the events that occur in a timeframe, it can be represented as:
\begin{equation}\label{3.1.3}
AS_{fore}=\sum\nolimits_{i=1}^{n} \sum\nolimits_{j=0}^{num_{i}} S_{fi}(event_{ij})
\end{equation}
Where $event_{ij}$ is the $j^{th}$ occurred event that belongs to foreground feature $f_i$, $n$ is total number of foreground features, $num_i$ is the number of occurred events for specific foreground feature $f_i$. The background anomaly score $AS_{back}$ computed by the model can be represented as:
\begin{equation}\label{3.1.4}
AS_{back}=\sum\nolimits_{i=1}^{m} \Big(S_{bi}(b_i) - \mu \times t_i\Big)
\end{equation}
Where $\mu$ is the damping factor which compensates for long periods of inactivity and allows for gradual decrease in the anomaly score, $t_i$ is the duration of inactivity, $m$ is total number of background features. The reference computation module is responsible for computing foreground reference $ref_{fore}$ and background reference $ref_{back}$ shown as follows:
\begin{equation}\label{3.1.5}
\begin{cases}
ref_{fore_i} = f_{ref}\Big(ref_{fore_{i-1}},AS_{fore_i} \Big)    \\
ref_{back_i} = f_{ref}\Big(ref_{back_{i-1}},AS_{back_i} \Big)    \\
\end{cases} 
\end{equation}
The foreground and background anomaly scores and respective references are then utilized together as input parameters of the trained ANFIS model to generate the real-time threat level. Thus, threat level of current user computed by the decision module can be represented as:
\begin{equation}\label{3.1.6}
threat \textunderscore level=ANFIS(p_{1},p_{2},...,p_{i}) 
\end{equation}
where in our case we have 4 inputs: $p_1=AS_{fore}$, $p_2=AS_{back}$, $p_3 = (AS_{fore}-ref_{fore})$, $p_4 = (AS_{back}-ref_{back})$. The computed $threat \textunderscore level$ is then used to make final authentication decisions.
\subsection{Features used in ANFIS system}
\label{sec3.2}
The proposed ANFIS-based authentication system utilizes representative features to model user profile. Also, these features should be easily obtainable and readily available from most mobile devices. Hence, the following sources are considered as features for our ANFIS-based authentication system. The experiments presented in this paper are based on these eight features for profile modeling. 

\begin{enumerate}[label=\arabic*),leftmargin=*,labelindent=5mm,labelsep=3mm,topsep=0pt]
\item Incoming SMS 
\item Outgoing SMS 
\item Incoming Call 
\item Outgoing Call 
\item Browser History 
\item WIFI History 
\item Screen Status 
\item Application History 
\end{enumerate}

These features are divided into foreground features(Screen status, app, outgoing sms/call) and background features(WIFI, incoming sms/call), and then employed within the authentication scheme to generate foreground and background anomaly scores (using equations 10 and 11) to quantify the authenticity of the current user. The score for each feature is calculated based on different conditions shown in Table~\ref{table_scf}. The score will reflect the user as more legitimate if more of the conditions are met. For example, when the user makes an outgoing call, the system will be triggered to check the following: 

\begin{enumerate}[label=$\circ$,leftmargin=*,labelindent=5mm,labelsep=3mm,topsep=0pt]
\item is the number in the user's contact list?
\item what is the position of the number in the ranking list (calculated using the ranking algorithms described in the next subsection)?
\item does the activity duration (in this case a call) fall within a given value? 
\end{enumerate}

Since an anomaly based scoring approach is utilized, more `legitimate' activities will lead to lower scores (i.e. more conditions in Table~\ref{table_scf} are met), while higher scores will indicate `illegitimate' or adversary activity (i.e. less of the conditions in Table~\ref{table_scf} are met). Next, we describe the generalized ranking algorithm which is used for most of the features within the ANFIS-based IA system.

\begin{table}[h]
\begin{center}
\caption{Conditions checked while computing anomaly scores for each of the monitored foreground/background features (activities).}
\label{table_scf} 
\begin{tabular}{|c||c|c|c|c|}
\hline
& RankingList & ContactList & Duration\\ 
\hline
Incoming/Outgoing call & \ding{51} & \ding{51} & \ding{51}\\
\hline
Incoming/Outgoing SMS & \ding{51} & \ding{51} &\\
\hline
WIFI & \ding{51} & & \ding{51}\\
\hline
Application history & \ding{51} & & \ding{51}\\
\hline
Browser history & \ding{51} & & \ding{51}\\
\hline
Screen On/Off & & & \ding{51}\\
\hline
\end{tabular}
\end{center}
\end{table}

\subsection{Generalized algorithm for maintaining a ranked list of occurred events within each feature category.}
\label{sec3.3}
To ensure that the most relevant and current items within each feature category are fully utilized, we developed a generalized algorithm for maintaining a ranked list. The main concept of ranking list is to quantify the importance of each item in the related feature category. For instance, if the user is using a specific application like Twitter, the system should be able to know if this is a frequently used application to the user, or how familiar this application is to the user compared with other applications. Our generalized algorithm is developed for this purpose. The algorithm is divided into two parts which are ranking list maintenance part and anomaly score computation part. The ranking list maintenance part is shown in Equation~\ref{3.3.1} to determine its ranking or index in the ranking list. 
\begin{equation}\label{3.3.1}
val_{ij}=
\begin{cases}
\alpha_{i}+\beta_{i}\times occ_{j}, & \text{if}\ item_{j} \notin rankList_{i} \\
val^{'}_{ij} + \beta_{i} - \lambda_{i} \times t_{j}, & \text{if}\ item_{j} \in rankList_{i} \\
\end{cases}
\end{equation}
Note that for any feature category $feature_{i}$, there is a ranking list $rankList_{i}$ where the $val_{ij}$ for each $item_{j}$ (that only belongs to that category) is stored. The value of each $item_{j}$ in its belonging feature category $feature_{i}$ is represented as $val_{ij}$, $val_{ij}$ is calculated and updated based on number of occurrence \textit{occ} and time \textit{t} (unit: hours) that has elapsed since last occurrence. The upper formula in Equation~\ref{3.3.1} is used to calculate the $val_{ij}$ if the $item_{j}$ is not in $rankList_{i}$ (event has not occurred before), otherwise lower formula is used to update $val_{ij}$ (item has existed and is stored in ranking list, $val^{'}_{ij}$ stores the same value as $val_{ij}$). Based on the given formula in Equation~\ref{3.3.1}, we can see that the higher the $val_{ij}$ is, the more relevant $item_{j}$ is to the user. If $val_{ij}$ is below $0$, then $item_{j}$ will be removed from the $rankList_{i}$. $\alpha$, $\beta$ and $\lambda$ can be adjusted based on the character of each feature and different user behaviour patterns. The scoring part of developed generalized scoring algorithm for ranking is defined in equation~\ref{3.3.2} and~\ref{3.3.3}.
\begin{equation}\label{3.3.2}
\epsilon = \frac{\mu'}{2}\times \left(\left(\frac{index-1}{len}\right)^2 +\left(\frac{vmax-val}{vmax-vmin}\right)^2\right)
\end{equation}
\begin{equation}\label{3.3.3}
    AS_{ranking}=
    \begin{cases}
      0, & \text{if }\ len=0 \\
      \frac{\mu}{2}, & \text{if }\ len=1,2 \text{ and } index = 0 \\
      0, & \text{if }\ len=1,2 \text{ and } index = 1 \\
      \mu, & \text{if }\ len>2 \text{ and } index = 0 \\
      \epsilon, & \text{if }\ len>2 \text{ and } index >= 1 \\
    \end{cases}
\end{equation}
where $AS_{ranking}$ is the anomaly score for ranking list while $\mu$ is the maximum value for $AS_{ranking}$. $val$ is the value of the item stored in ranking list and $index$ is the ranking position of the item (item with the highest value is indexed by 1, $index=0$ indicates the item is not in top list). $len$ is the length of ranking list, $vmax$ and $vmin$ is the item with the highest and the lowest value in ranking list respectively. $\mu'$ represents the adjusted maximum value for $AS_{ranking}$ which is shown in the equation below.
\begin{align}\label{3.3.4}
\begin{split}
\mu' = \frac{\mu}{1+e^{-c(len -b)}}
\end{split}
\end{align}
where $c$ and $b$ is the parameter for the sigmoid function. The main functionality of equation~\ref{3.3.4} is to reduce the maximum value for $AS_{ranking}$ when the length of ranking list is less than the stipulated amount. This can be adjusted based on the behavior of different users. To summarize, the developed algorithm for ranking is capable of self-updating adaptively in concert with the user behaviour drift by maintaining a cache of the most relevant items for each feature.

\subsection{Time window-based profiling}
\label{sec3.4}
Time window-based profiling technique is designed with the aim of producing a more granular user model. To achieve this objective, screen status is utilized to capture every activity performed by the user. That is, whenever user presence is detected, the system will track all the foreground activities performed by the user and compute the foreground anomaly score $AS_{fore}$ incrementally until screen off is detected. The system will then proceed to reference computation and ANFIS module to update $ref_{fore}$, $ref_{back}$ and generate authentication result. Whenever $AS_{fore}$ is recalculated and the system is in deployment phase, the authentication will be performed based on new $AS_{fore}$ and current $AS_{back}$ with damping factor considered. The occurrences of background activities will only update the score of their related features and no authentication will be performed. The main functionality of timer (see Fig.~\ref{fig_sbd}) is to control the maximum time period that the system has to wait to perform authentication when user is active, this forces the system to always provide security within stipulated timeframe.
\subsection{Adaptive reference computation}
\label{sec3.5}
As mentioned earlier, a reference is utilized as an ANFIS input variable in tandem with foreground and background anomaly score. The reference is calculated in a way that indicates the distributions or patterns of anomaly scores. Therefore the system can decide if the current behaviour deviates too much from previously perceived patterns by calculating the difference between anomaly score and reference. Thus, to generate dynamic reference which is adaptive to user behaviour drift, we choose exponentially weighted moving average (EWMA) and standard deviation (SD) as the main parts of our reference computation scheme. Previous works such as~\cite{bib20} where EWMA is applied to admission control and buffer management~\cite{bib21} have illustrated the performance and efficiency of the algorithm. In EWMA, the coefficient \(\alpha\) represents the degree of weighting, such that a higher \(\alpha\) discounts older observations (thresholds) faster as shown in equation~\ref{3.3.1}.
\begin{align}\label{3.5.1}
\begin{split}
reference_{t}=\alpha\times AS+(1-\alpha)\times reference_{t-1}
\end{split}
\end{align}
Therefore, low value \(\alpha\) is considered appropriate in our case so that the new computed threshold changes steadily and gradually converges with the current aggregate score. Thus, the value of \(\alpha\) is set at 0.2 throughout the experiments. The mechanism of our EWMA\textunderscore SD\textunderscore BLOCK (ESB) reference computation scheme is illustrated in Algorithm~\ref{EWMAsdblock}.

\begin{algorithm2e}
 \caption{ESB reference computation algorithm}
 \KwData{newly computed foreground/background anomaly score \textit{AS}, ESB preparation length \textit{len} in days, block size \textit{b}}
 \KwResult{foreground/background reference \textit{ref}}
 
 OnApplicationInstalled:\\
 $ESBPreparation.isFinished \longleftarrow false$;\hspace{3mm}\textit{b} $\longleftarrow$ 7;\hspace{3mm}i $\longleftarrow$ 0\;
 \If{new AS is detected}{
	\uIf{ESBPreparationDuration \(<\) len \textbf{and} ESBPreparation.isFinished \(=\) false}{
		\textbf{continue}\;}
	\uElseIf{ESBPreparationDuration \(>=\) len \textbf{and} ESBPreparation.isFinished \(=\) false}{
		\tcc*[h]{$mean$ and $standard\_deviation$ for all the $ASs$ within preparation duration, $ref$ is used for next block of $ASs$(size $b$)}\par
		$ref$ $\longleftarrow$ $mean - standard\textunderscore deviation$\;  
		$ESBPreparation.isFinished \longleftarrow true$\;}
	\ElseIf{ESBPreparation.isFinished \(=\) true}{
		\tcc*[h]{preparation ends,computed $ref$ can now be used as ANFIS inputs}\par
		\uIf{i \(<\) b}{
			$sum += AS;$\hspace{3mm}$i++$\;}
		\ElseIf{i \(=\) b}{
			$ave \longleftarrow sum/b$;\hspace{3mm}  $sum \longleftarrow AS$;\hspace{3mm}  $i \longleftarrow 1$\;
			\tcc*[h]{newly computed $ref$ is used for next block}\par
			$ref_{t} \longleftarrow \alpha\times ave+(1-\alpha)\times ref_{t-1}$ \;}		
    }
 }
\vspace{0.5cm}
\label{EWMAsdblock}
\end{algorithm2e}

\subsection{The ANFIS classifier}
\label{sec3.6}
The main functionality of ANFIS in our authentication system is to train a specific fuzzy inference model for each user based upon their behaviour through ANFIS learning process, then to provide real-time authentication based upon given input variables when training is finished. As mentioned before, we utilize foreground/background anomaly score and computed reference in concert as ANFIS inputs. Thus, the selected ANFIS input variables are shown as follows(see equation~\ref{3.1.6}):
\begin{enumerate}[label=$\circ$,leftmargin=*,labelindent=5mm,labelsep=3mm,topsep=0pt]
\item foreground anomaly score($AS_{fore}$)
\item the difference of foreground anomaly score and reference($ARD_{fore}$)
\item background anomaly score($AS_{back}$)
\item the difference of background anomaly score and reference($ARD_{back}$)
\end{enumerate}
while output data is classified into three categories:
\begin{enumerate}[label=$\circ$,leftmargin=*,labelindent=5mm,labelsep=3mm,topsep=0pt]
\item output = 1 indicates legitimate user
\item output = 0 indicates suspicious user
\item output = -1 indicates adversary
\end{enumerate} 

\section{Experiments and Evaluation}
\label{sec4}
This section presents the experimental results and evaluation of the proposed ANFIS-based implicit authentication system. The performance between implicit authentication systems with and without implementation of ANFIS will be compared. An Android application is developed to extract the data based on the selected features. In order to automate the evaluation process and facilitate future research, the scoring module, reference computation module and ANFIS module are implemented by using Python scripts to emulate the logic of the ANFIS-based authentication system on mobiles devices. Five user data sets including respective adversary cases are collected over several weeks as shown in Table~\ref{table_userdata}. In order to test the efficiency and resilience of the proposed authentication system on dealing with both normal and complex adversary scenarios, two adversary cases `uninformed attacker' and `informed attacker' were investigated for every user to emulate real-life device theft.

\begin{table}[h]
\begin{center}
\caption{Information of collected participants' data  }
\label{table_userdata}
\begin{tabular}{l | *{2}{c}} \hline\hline
User ID & Number of event logs & Collection time period \\ \hline
\\[-1em]
User 1  &   11107             &      2016/06/29 -- 2016/07/30      \\ \hline
\\[-1em]
User 2  &   23628             &      2016/06/29 -- 2016/09/18       \\ \hline
\\[-1em]
User 3  &   2847             &       2016/06/30 -- 2016/09/25     \\ \hline
\\[-1em]
User 4  &   9319             &       2016/09/18 -- 2016/12/11    \\ \hline
\\[-1em]
User 5  &   17945             &      2016/09/21 -- 2016/12/11     \\ \hline\hline

\end{tabular}
\end{center}
\end{table}
 
\textbf{\textit{Uninformed attacker}}: The attacker takes the device and interacts with it for several hours. The attacker has no knowledge of how the user behaves routinely. E.g. the attacker might connect to a WIFI previously unknown to the device, use unusual applications which are different from the owner's.

\textbf{\textit{Informed attacker}}: The attacker takes the device and interacts with it for several hours. In this circumstance, the attacker was provided with some additional information regarding the user behavior. E.g. a list of applications and WIFI access points that the owner most frequently used, whom the owner mostly contacted with, which websites the owner usually browses.

In order to evaluate the ANFIS model, the data collected from each user is split into training and testing portions. The training part is labelled with the correct output while the testing part is unlabelled. After training, the model is evaluated on the test part and the correct predictions made are computed to give the final performance metrics.

User recognition rate and elapsed time are chosen as metrics to evaluate accuracy and adversary detection performance. As illustrated in Table~\ref{table_ANFISornot}, with the adoption of ANFIS, the user recognition rate is increased from an average of 90\% to an average of 95\% compared with a non-ANFIS-based system. For the adversary performance, ANFIS is slightly outperformed by non-ANFIS, which is tolerable considering the improvement that was achieved by introducing the ANFIS system.

Table~\ref{table_ANFISresult} and Table~\ref{table_nonANFISresult} presents the experimental results of five users to evaluate the performance of ANFIS-based and non-ANFIS implicit authentication system respectively. It shows that ANFIS is suitable and efficient for our proposed implicit authentication system with 95\% average accuracy and adversary detection capability within a few minutes. Besides, ANFIS is able to automatically build a tailored fuzzy inference model for every user based on their training data. It doesn't require prior understanding of user behaviour pattern and eradicates the need of cumbersome tuning tasks like in traditional fuzzy-logic method. Overall, our proposed novel ANFIS-based implicit authentication system has proven to be promising at providing intelligent mobile access control.

\begin{table}[h]
\begin{center}
\caption{Experimental results from ANFIS method}
\label{table_ANFISresult}
\begin{tabular}{l | *{3}{c}} \hline \hline
\raisebox{1.2ex}{User ID} & \shortstack{User Recognition\\ Rate} & \shortstack{Elapsed Time\\(informed adversary)} & \shortstack{Elapsed Time\\(uninformed adversary)}\\ \hline
\\[-1em]
User 1  &   93.01\%             &  9.88    min   &  9.88  min  \\ \hline
\\[-1em]
User 2  &   96.55\%             &  16.88    min    &  16.88  min  \\ \hline
\\[-1em]
User 3  &   98.21\%             &  3.88     min    &  3.88  min   \\ \hline
\\[-1em]
User 4  &   96.64\%             &   4.88    min    &  4.88  min   \\ \hline
\\[-1em]
User 5  &   90.36\%             &   11.37   min    &  7.38  min   \\ \hline
\\[-1em]
Average  &   94.95\%             &  9.38   min    &  8.58  min    \\ \hline\hline
\end{tabular}
\end{center}
\end{table}

\vspace{-0.3cm}

\begin{table}[h]
\begin{center}
\caption{Experimental results from non-ANFIS method}
\label{table_nonANFISresult}
\begin{tabular}{l | *{3}{c}} \hline \hline
\raisebox{1.2ex}{User ID} & \shortstack{User Recognition\\ Rate} & \shortstack{Elapsed Time\\(informed adversary)} & \shortstack{Elapsed Time\\(uninformed adversary)}\\ \hline
\\[-1em]
User 1  &   86.81\%             &  12.88    min   &  3.88  min  \\ \hline
\\[-1em]
User 2  &   90.91\%             &  13.88    min    &  13.88  min  \\ \hline
\\[-1em]
User 3  &   98.08\%             &  6.88     min    &  6.88  min   \\ \hline
\\[-1em]
User 4  &   88.84\%             &   4.88    min    &  4.88  min   \\ \hline
\\[-1em]
User 5  &   87.36\%             &   11.37   min    &  7.38  min   \\ \hline
\\[-1em]
Average  &   90.40\%             &  9.98   min    &  7.38  min    \\ \hline\hline
\end{tabular}
\end{center}
\end{table}

\vspace{-0.3cm}

\begin{table}[h]
\begin{center}
\caption{Performance comparison with and without ANFIS}
\label{table_ANFISornot}
\begin{threeparttable}
\begin{tabular}{l | *{2}{c}} \hline\hline
Average of 5 users& \shortstack{Non-ANFIS} & \shortstack{ANFIS} \\ \hline
User Recognition Rate & 90.40\%  & 94.95\% \\ 
Elapsed Time(uninformed adversary) & 7.38 mins & 8.58 mins \\ 
Elapsed Time(informed adversary) & 9.98 mins  & 9.38 mins \\ \hline\hline

\end{tabular}
\begin{tablenotes}
  \item[*] Elapsed Time: time elapsed after the attack starts
\end{tablenotes}
\end{threeparttable}
\end{center}
\end{table}

\section{Related Work}
\label{sec5}
Research regarding implicit or transparent authentication has been increasing in the last few years. Implicit authentication was firstly proposed in~\cite{bib13} which records the user's information including location, phone calls, SMS and browser events. Its scoring system is based on the probability density function. However, the issue of tackling behaviour drift is not addressed. In contrast with~\cite{bib13}, the work of Kayacik et al.~\cite{bib8} presents a better performance regarding adaptiveness and transparency. Their scheme utilizes probability density function to model user behaviour and is able to activate retraining phase when behaviour drift is detected. However, this scheme samples the sensor data including GPS frequently which might not be power efficient.

The authors in~\cite{bib10} proposed a novel behaviour profiling framework to authenticate users through analyzing their application usage patterns. The proposed framework presents an more advanced and optimized utilization of app usage pattern than the way it is utilized in~\cite{bib5,bib8}.  The framework measures the authenticity of a user based on a number of consecutive abnormal application activities. Their method also maintains a dynamic profile which contains 7/10/14 days of users' most recent activities, updated on a daily basis. However, the research is conducted on MIT Reality dataset that was created in 2004 with limited number of applications. Similar to~\cite{bib10}, authors in~\cite{bib9} proposed another application-centric implicit authentication. Instead of extracting location and time from the application,~\cite{bib9} records the touchscreen patterns of the user to build the behavioural profile.

Another context profiling framework is proposed in~\cite{bib6} which utilized context variables including GPS readings, surrounding devices via both WIFI and Bluetooth. This framework firstly detects personal contexts of interest (CoIs) and then infer user trust level based on device familiarity and context familiarity. However, this framework is not able to adaptively update CoIs when a user's routine regarding location changes significantly. Later as a follow-up study, an improved version based on~\cite{bib10} called ConXsense is proposed in~\cite{bib15} to protect against device misuse. The identification of CoIs is optimized by using stay points and stay regions to identify CoIs based on GPS and WIFI. ConXsense is designed not to authenticate the user, but to provide smart device locking system. With the similar aim, a context-aware scalable authentication scheme is proposed in~\cite{bib11} which is designed to reduce active authentication on the mobile device by inferring the safety level of the context. 

Different from all the previous works, this paper investigates an ANFIS-based system for implicit authentication on mobile devices and uses real data obtained from users over several weeks for the evaluation of the system.
\section{Conclusion}
\label{sec6}
%
%

This paper proposed and investigated an ANFIS-based system for continuous implicit authentication for mobile devices. The system was evaluated using data collected over several weeks from mobile device users where an Android app continuously monitored 8 different activities (features) to collect data for our the experiments presented in this paper. Through its self-learning capability, an ANFIS-based system provides tailored fuzzy inference system for every user based on their collected behavioural data, thereby eliminating the need for cumbersome manual tuning required in traditional fuzzy based systems. The experimental results presented showed that the ANFIS based system achieved up to 95\% user recognition rate on average compared to a non-ANFIS system which recorded 90\% user recognition rate on average. In future work we intend to undertake further experiments by complementing our real user dataset with simulated data in order to enable more in-depth evaluation on a larger user space.






\begin{thebibliography}{1}
\bibitem{bib1}
`One in every six users suffer loss or theft of mobile devices', http://www.kaspersky.com/about/news/press/2013/one-in-every-six-users-suffer-loss-or-theft-of-mobile-devices, accessed 21 December 2016

\bibitem{bib2}
G.X. Ye, T.Z. Tang, D.Y. Fang, X.J. Chen, K.I. Kim, B. Taylor and Z. Wang, ``Cracking Android pattern lock in five attempts.'' In: The Network and Distributed System Security Symposium 2017 (NDSS'17).

\bibitem{bib3}
A.N. Kataria, D.M. Adhyaru, A.K. Sharma and T.H. Zaveri, ``A survey of automated biometric authentication
techniques.'' 2013 Nirma University International Conference on Engineering (NUiCONE), Ahmedabad, 2013, pp. 1--6

\bibitem{bib4}
H. Crawford and K. Renaud, ``Understanding user perceptions of transparent authentication on a mobile device.'' Journal of Trust Management, 2014, \textbf{1}, (1), pp. 1--28

\bibitem{bib5}
L. Fridman, S. Weber, R. Greenstadt and M. Kam,
``Active authentication on mobile devices via stylometry, application usage, web browsing, and GPS location.'' arXiv preprint arXiv: 1503, 08479, 2015, pp. 1--10

\bibitem{bib6}
A. Gupta, M. Miettinen, N. Asokan and M. Nagy, ``Intuitive security policy configuration in mobile devices using context profiling.'' Privacy, Security, Risk and Trust (PASSAT), 2012 International Conference on and 2012 International Confernece on Social Computing (SocialCom), Amsterdam, 2012, pp. 471--480

\bibitem{bib7}
E. Hayashi, S. Das, S. Amini, J.
J Hong, and I. Oakley, ``Casa: context-
aware scalable authentication.'' In n Proceedings of the Ninth Symposium
on Usable Privacy and Security, pp.
p 3-13. ACM, 2013.

\bibitem{bib8}
H. G. Kayacik, M. Just, L. Baill lie, D. Aspinall, and N. Micallef, ``Data driven authentication: On the effectiveness of user behaviour modelling with mobile device sensors.'' In Proceedings of the Third Workshop on Mobile Security Technologies (MoST), 2014

\bibitem{bib9}
H. Khan and U. Hengartner, ``Towards application-centric
implicit authentication on smartphones.'' In Proceedings of the 15th Workshop on Mobile Computing Systems and Applications, ACM, New York, 2014, pp. 10:1--10:6

\bibitem{bib10}
F. Li, N. Clarke, M. Papadaki and P. Dowland, ``Active authentication for mobile devices utilising behaviour profiling.'' International journal of information security, Springer Berlin Heidelberg, 2014, \textbf{13}, (3), pp. 229--244.

\bibitem{bib11}
M. Miettinen, S. Heuser, W. Kronz, A. Sadeghi and N. Asokan, ``Conxsense - context sensing for adaptive usable access control.'' In Proceedings of the 9th ACM symposium on information, computer and communications security, 2014, pp. 293--304


\bibitem{bib12}
O. Riva, C. Qin, K. Strauss and D. Lymberopoulos, ``Progressive authentication: Deciding when to authenticate
on mobile phones.'' In Presented as part of the 21st USENIX Security Symposium, 2012, pp. 301--316

\bibitem{bib13}
E. Shi, Y. Niu, M. Jakobsson and R. Chow, ``Implicit authentication through learning user behavior.'' Information Security, Springer Berlin Heidelberg, 2010, \textbf{6531}, pp. 99-113

\bibitem{bib14}
F. Yao, S.Y. Yerima, B. Kang and S. Sezer, ``Event-driven
implicit authentication for mobile access control.'' 2015 9th International Conference on Next Generation Mobile Applications, Services and Technologies, Cambridge, 2015, pp. 248--255

\bibitem{bib15}
F. Yao, S.Y. Yerima, B. Kang and S. Sezer, ``Fuzzy logic-
based implicit authentication for mobile access control.'' SAI Computing Conference (SAI), IEEE, London, 2016, pp. 968--975.

\bibitem{bib16}
J.S.R. Jang, ``ANFIS: adaptive-network-based fuzzy inference system.'' in IEEE Transactions on Systems, Man, and Cybernetics, 1993, \textbf{23}, (3), pp. 665--685

\bibitem{bib17}
S.A. Khan, M.D. Equbal and T. Islam, ``A comprehensive comparative study of DGA based transformer fault diagnosis using fuzzy logic and ANFIS models.'' in IEEE Transactions on Dielectrics and Electrical Insulation, 2015, \textbf{22}, (1), pp. 590--596.

\bibitem{bib18}
Suraj, R.K. Sinha and S. Ghosh, ``Jaya Based ANFIS for Monitoring of Two Class Motor Imagery Task.'' in IEEE Access, 2016, \textbf{4}, pp. 9273--9282.

\bibitem{bib19}
Y. Liu, W. Zhang, and Y. Zhang, ``Dynamic Neuro-Fuzzy-Based Human Intelligence Modeling and Control in GTAW.'' in IEEE Transactions on Automation Science and Engineering, 2015, \textbf{12}, (1), pp. 324--335.

\bibitem{bib20}
S.Y. Yerima, ``Implementation and evaluation of
measurement-based admission control schemes within
a converged networks qos management framework.'' International Journal of Computer Networks and Communications, IJCNC, 2011, \textbf{3}, (4), ISSN: 0974-9322[Online], 0975-2293[print]

\bibitem{bib21}
S.Y. Yerima and K. Al-Begain, ``Novel radio link buffer management schemes for end-user multi-class traffic in high
speed packet access networks.'' Wireless Personal Communications, 2011, \textbf{61}, (2), pp. 349--382

%
%
%
%
%
%
%
%
%
%
%
%
%
%
%
%
%
%
%
%


\end{thebibliography}
%

\end{document}